\documentclass[journal=ancac3]{achemso}

\usepackage[version=3]{mhchem}  
\usepackage{graphicx}		
\usepackage{epstopdf}
\usepackage{dcolumn}		
\usepackage{bm}			
\usepackage{subfigure}
\usepackage{verbatim}		
\usepackage{color}
\usepackage{fixltx2e}
\usepackage{caption}
\usepackage{threeparttable}

\newcommand{\cs}{C$_{\textrm{60}}$\ }

\author{Mohammad Koleini}
\email{koleini.m@gmail.com}
\affiliation[University of Bremen]
	{Hybrid Materials Interfaces Group, Faculty of Production Engineering
 	and Bremen Center for Computational Materials Science,
	University of Bremen, 28359 Bremen, Germany}
\author{Mads Brandbyge}
\email{mads.brandbyge@nanotech.dtu.dk}
\affiliation[Technical University of Denmark]
	{DTU Nanotech, Department of Micro and Nanotechnology, Technical University of Denmark,
	\O rsteds Plads, Building 345E, DK-2800 Kongens Lyngby, Denmark}

\title[\texttt{achemso} ]
{Strong Spin-Filtering and Spin-Valve Effects in a Molecular V-C$_{\textrm{60}}$-V Contact}

\begin{document}

\begin{abstract}
Motivated by the recent achievements in manipulation of \cs molecules in STM experiments,
we study theoretically the structure and electronic properties of a \cs molecule in an STM-tunneljunction
with a magnetic tip and magnetic adatom on a Cu(111) surface from first-principle calculations.
For the case of V tip/adatom, we demonstrate how spin-coupling between the magnetic V atoms mediated by
the \cs can be observed in the electronic transport, which display a strong spin-filtering effect,
allowing mainly majority-spin electrons to pass(>95\%).
Moreover, we find a significant change in the conductance between parallel and anti-parallel spin polarizations
in the junction (86\%) which suggests that STM experiments should be able to characterize the magnetism and spin-coupling
for these systems.
\end{abstract}

%\keywords{Spin Transport, Fullerene, Scanning Tunneling Microscope, Molecular Spintronics,}

\section{INTRODUCTION}

Organic materials typically offer small spin-orbit and hyperfine interactions, which are prerequisites for spintronic applications, due
to allowing long spin lifetimes.
Thus there is a great interest in organic building blocks for molecular spintronics\cite{XiWuVa.2004,RoGaBa.2005,KoWaBl.2011,ISI:000275868700022},
and a thorough understanding of spin transport and magnetism in these systems is called for. It is therefore important to establish model molecular spintronic systems where spin-transport and magnetic interactions can be examined experimentally.

Recently, it has been demonstrated in low temperature scanning tunneling microscopy (STM) experiments how \cs molecules can be picked up by the STM-tip, and subsequently controllably used to contact structures such as adatoms, clusters, and molecules placed on a substrate
surface\cite{ScFrBr.2009,BeKrNe.2010a,ScFrAr.2011}.
The \cs molecule is considered as an attractive anchoring group for molecular electronics due to its mechanical
robustness~\cite{martin2008fullerene}. Moreover, the lowest unoccupied molecular orbital (LUMO) of \cs is close to the Fermi level of ferromagnetic elements, which makes spin injection relatively easy~\cite{braun2009energy}, and makes \cs a promising building block in molecular spintronics.

The high symmetry of the \cs allows detailed characterization of the bonding geometries in STM. In particular,
one can determine which part of \cs is pointing towards the tip/surface prior to contact formation, and also afterward, while the \cs is placed on the tip\cite{ScFrAr.2011}. Subsequently, it is possible to investigate the interactions between tip and sample via electronic transport measurements, as the tip and and sample are brought into contact.
STM is a powerful tool for investigating spin-transport in magnetic
nanostructures~\cite{Bo.2003,PhysRevLett.65.247,hirjibehedin2006spin,ISI:000286768600002,ISI:000280416200031}.
Direct magnetic interaction between STM tip and magnetic materials on a substrate has been studied in number of
works\cite{tao2009switching,PoFeBa.2011,BrGa.2009}, and STM has been used to probe spin in organic
molecules~\cite{atodiresei2010design}.
In the case of a magnetic tip and magnetic surfaces, this method may be used to study spin-transport and interactions though organic molecular systems bound to the surface and gain insight into single-molecular magnetic properties.

Among organic compounds, carbonic rings when combined with transition metals show versatile combinations
for molecular spintronics~\cite{crabtree2009organometallic}. The interaction of $\pi$ electron system of such rings
with $d$ orbitals of the transition metals, is a key to electron and
subsequently spin transport. One example of such systems is presented in a theoretical study, where calculations have been
used to examine spin transport in a benzene-Co system on a Cu(001) surface contacted
by a Cr tip~\cite{tao2008manipulating}. Further on, the magnetic properties and spin transport have been calculated
for organometallic ``multidecker'' wires where magnetic atoms are sandwiched between organic parts\cite{WaCaWa.2008}.
Multidecker systems involving vanadium appear very promising due to their half-metallic behavior
resulting in high spin-polarization of the transport\cite{MaBaMe.2006,KoPaBr.2007,WaCaWa.2008}.
Interstingly, due to the different symmetries of the \cs, it is possible to vary the electronic and magnetic properties depending
on whether \emph{pentagon}, \emph{hexagon} or \emph{edge} sites of \cs is in contact with magnetic ligand atoms.

Here, we employ first principles calculations to predict the spin transport through spintronic model system consisting of a \cs molecule contacted by
magnetic atoms in a STM setup. In particular, we predict that vanadium is a magnetic material where pronounces spin-filtering and spin-valve effects in STM experiments.

\section{SYSTEM AND METHOD}

\begin{figure}[htbp]
   \includegraphics[width=0.4\columnwidth]{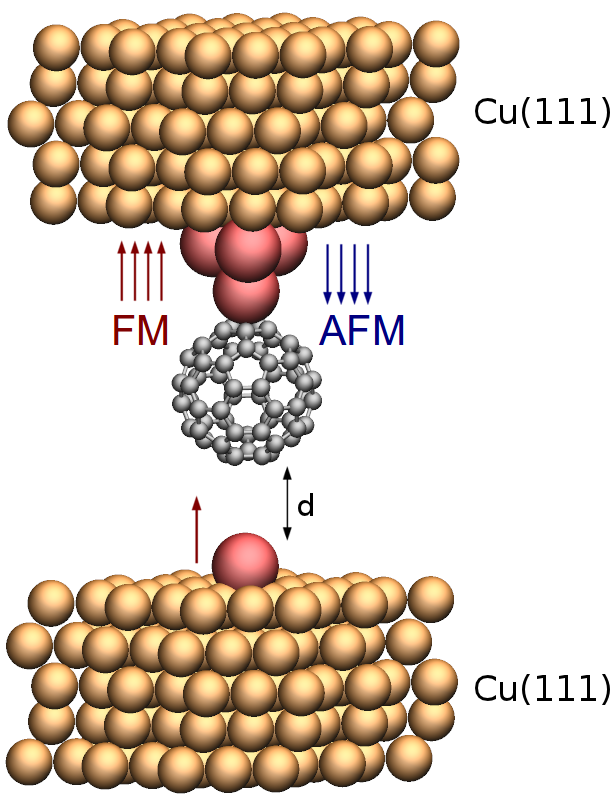}
   \caption{STM system picking up \cs with a magnetic tip, approaching a magnetic adatom on the nonmagnetic copper surface.
	The orange, red and gray spheres depict copper, vanadium and carbon atoms respectively.}
   \label{fig:fig1}
\end{figure}

In order to mimic a concrete STM experiment, we investigate the specific setup shown in~\ref{fig:fig1}.
The bulk regions of the contacts (i.e. STM tip and substrate) have been chosen to be non-magnetic copper,
which has previously been employed in manipulation experiments\cite{ScFrBr.2009,BeKrNe.2010a,ScFrAr.2011}.
We imagine that magnetic atoms are deposited on the Cu surface prior to deposition\cite{weber1997magnetism} of the \cs molecules, and that the tip-electrode is prepared prior to the contact by either indenting a Cu tip into a cluster of these atoms in order to cover its outermost part with these, or by creating the tip from from the bulk magnetic material.
We model the outermost part of the tip by a pyramid-like structure consisting of four atoms on one electrode,
simulating the magnetic STM tip which is used to pick up \cs for subsequent contact formation to an isolated adatom on the copper (111) substrate. In the following we will show that the case of vanadium is remarkable.

We used spin-polarized pseudopotential DFT calculations with the {\sc SIESTA} code~\cite{Soler2002}.
Electronic structures have been calculated within a GGA-PBE approximation to the exchange and correlation
functional~\cite{Perdew1996}. Double-$\zeta$ polarized basis sets with grid cutoff of 250 Ry have been used.
Spin polarized transport was subsequently calculated using the non-equilibrium Green's function (NEGF)
formalism\cite{Brandbyge2002} in the limit of zero voltage.
In order to eliminate basis set superposition error (BSSE) present in methods with
atomic orbitals basis sets, we have used plane-wave (PW) formalism as implemented
in Ref.~\cite{Giannozzi2009} for total energy calculations. In these calculations, we have been using
ultrasoft pseudopotentials, with 30/300 Ry cutoff for wavefunction/charge density.

To guide us in our search for interesting magnetic metals to contact the \cs, and to select relevant bonding sites,
we first performed calculations of the binding energy of a single adatom from the first row transition metals (from vanadium to copper) with \cs. Here our focus, besides achieving high magnetization, is to obtain a high binding energy in order to achieve a stable contact.
The results are summarized in~\ref{tab:ResSum} for different sites on the \cs molecule.
One observes that nickel has the strongest binding energy, but with zero total magnetization (M$_T$),
and thus is probably not interesting for investigations of spin-transport. On the other hand, chromium enjoys
the largest M$_T$, due to its largest unpaired electronic configuration [Ar]3d$^5$4s$^1$,
but with the least binding energy strength.
It has already been shown that copper STM tips can pick up \cs~\cite{ScFrBr.2009}. Noting that the maximum of
binding energy for copper to \cs is $\sim$ 0.8~eV, we conclude that the same action should
be possible with V with a decent M\textsubscript{T}.

\begin{table}[htbp]
  \begin{center}
    \captionsetup{singlelinecheck=off}
    \caption{Binding energy E\textsubscript{b} = E\textsubscript{C\textsubscript{60}+adatom}
	- E\textsubscript{C\textsubscript{60}} - E\textsubscript{adatom} and total magnetization
M\textsubscript{T} = $ \int_v (n_{up} - n_{down})\,dr^3 $ per unit cell.
The distance between the adatom and \cs has been optimized.}
    \label{tab:ResSum}
    \setlength{\tabcolsep}{8pt}
    \setlength{\extrarowheight}{2.5pt}
    \begin{tabular}{lcccccccc}
        \hline
        \hline
         adatom & 56 (eV) & 66 (eV) & hexagon (eV) & pentagon (eV) & top (eV) & M\textsubscript{T} \\
        \hline
	V  & -1.13 & -1.25 & -0.93 & -1.40 & -1.06  &	5.00\\
	Cr & -0.58 & -0.46 &  1.77 &  2.74 & -0.49  &	6.00\\
	Mn & -0.39 & -0.58 & -0.25 & -0.33 & -0.40  &	5.00\\
	Fe & -0.83 &  0.08 & -0.87 & -0.55 & -0.75  &	4.00\\
	Co & -0.90 & -1.22 & -0.87 & -0.88 & -0.80  &	1.00\\
	Ni & -1.48 & -1.64 & -1.24 & -1.44 & -1.26  &	0.00\\
	Cu & -0.79 & -0.60 & -0.08 & -0.34 & -0.74  &	0.00\\
	\hline
	\hline
    \end{tabular}
  \end{center}
\begin{tablenotes}\footnotesize
\item[1] 66 refers to sites above edges shared between two hexagons, 56 referes to sites above edges shared between
a hexagon and a pentagon and top means top of a carbon atom.
\end{tablenotes}
\end{table}

\begin{figure}[htbp]
   \includegraphics[width=\columnwidth]{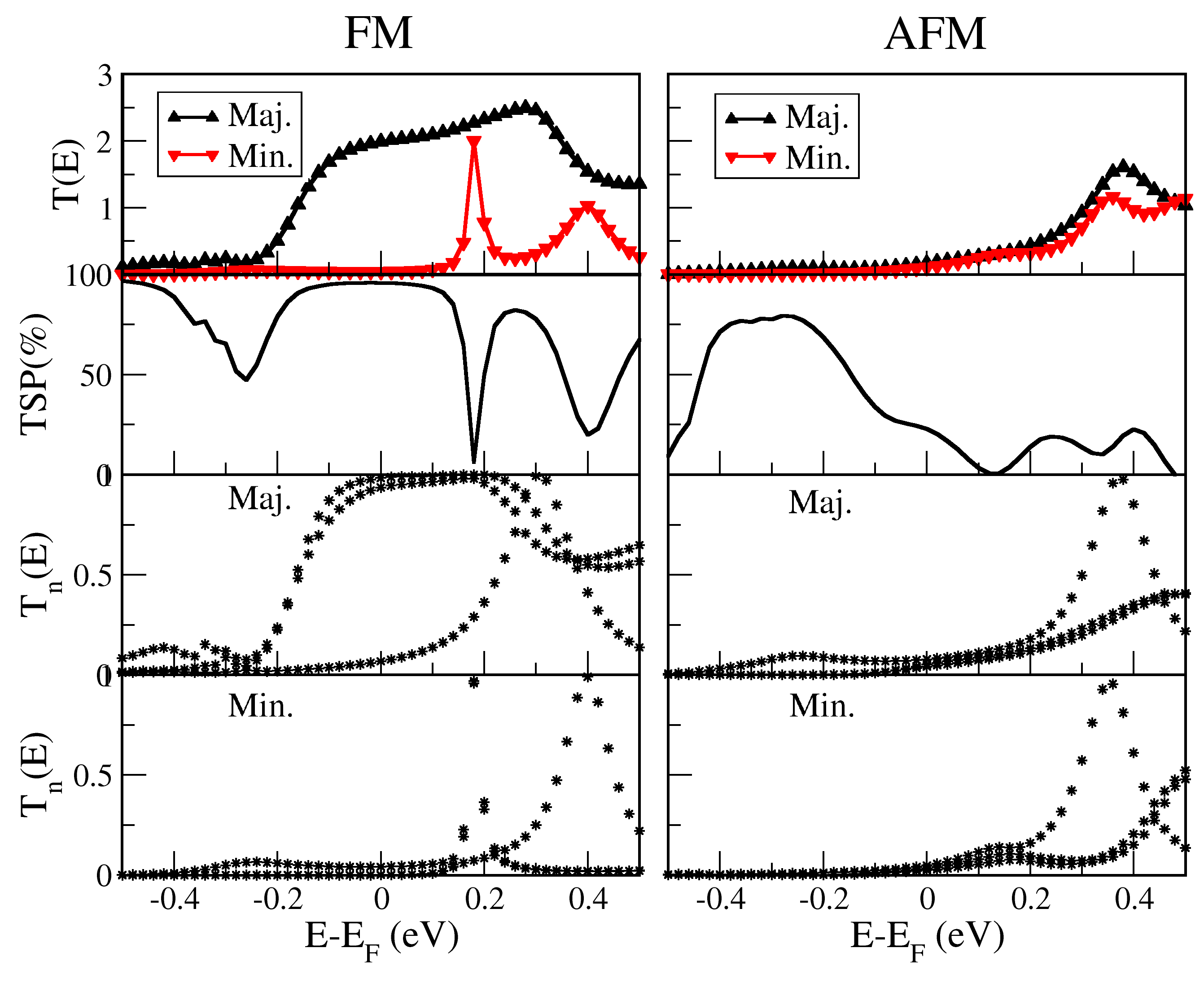}
   \caption{Transmission spectra for FM and AFM arrangements. The first row shows spin-resolved
	transmission spectra for each arrangement. The second row demonstrates the corresponding
	transmission spin polarization as defined in the text. The third and forth rows show transmission
	eigenvalues for three channels in each spin manifold.}
   \label{fig:fig2}
\end{figure}

Based on the data of the simple guiding calculations we subsequently chose vanadium in the full \cs-contact simulations on the system depicted in~\ref{fig:fig1}.
In the following we demonstrate that this choice for contact material in an STM setup indeed is successful in achieving a good spin-filter and spin-valve performance. Here, ferromagnetic (FM) and anti-ferromagnetic (AFM) spin alignment between atoms of tip and adatom, have been considered. The site on the \cs with highest binding energy for V adatom is $\eta^5$, roughly over the center of a pentagon of a \cs, and due to the symmetric structure of the \cs, this site is contacted both by the tip and the adatom.
We find a binding energy of the \cs to the V-tip (upper part in~\ref{fig:fig1}) is 1.3~eV, while binding of \cs to the adatom on the Cu substrate (lower part in~\ref{fig:fig1}) is 1.1~eV.
The spin-resolved transmissions for the FM and AFM cases are shown in~\ref{fig:fig2}. We first focus on the highly conducting contact configuration where the 
atomic structure of the \cs along with vanadium atoms and first copper layers of both sides have been relaxed to the force threshold of 0.05~eV/\AA.
We also show the transmission spin polarization (TSP), defined as
\begin{equation}
TSP=\frac{T_{maj}-T_{min}}{T_{maj}+T_{min}}\,,
\label{eq.TSP}
\end{equation}
and channel decomposed transmission values in~\ref{fig:fig2}.
Here, we first point out a remarkable {\em spin-filtering} effect in the FM arrangement,
whereby two almost open channels conduct in the vicinity of the Fermi level for the majority
spin component, while the minority channels are almost closed.
For the minority spin component the resonance peaks at
$\sim$ 0.2 and 0.4~eV produce dips in the corresponding TSP curve, however, these will only be of importance for a voltage
bias comparable to these energies. Transmission eigenvalues of first three dominant channels are shown in the third and forth
panels, that clearly show two distinct channels for the FM-majority spin channels.
Furthermore, it is striking that the channels in the AFM configuration are almost closed, except for small resonance peaks at $\sim$ 0.35~eV, which again only will come into play for higher voltages.

\begin{figure}[htbp]
   \includegraphics[width=0.7\columnwidth]{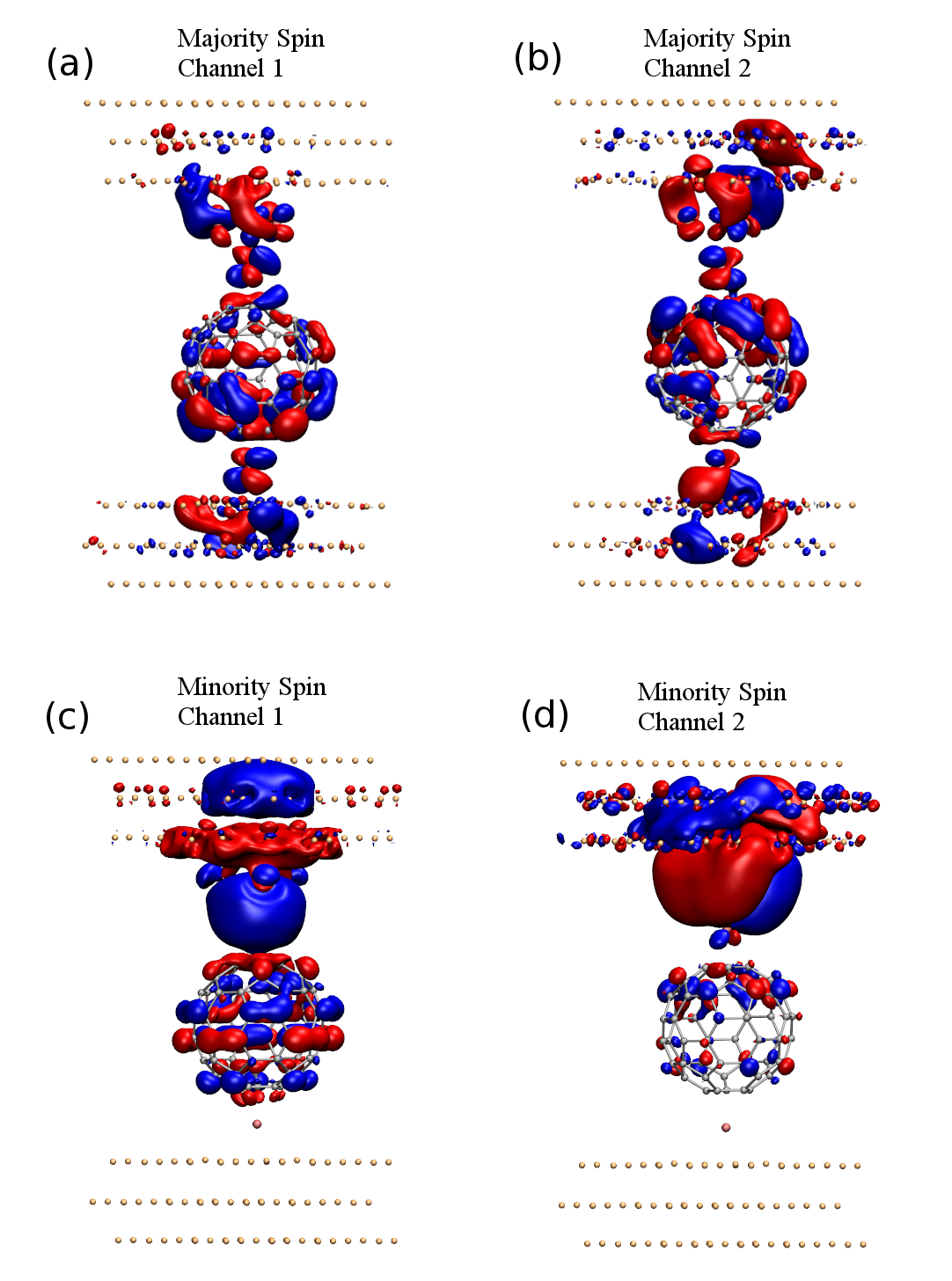}
   \caption{Scattering states at $E=E_F$ of first two dominant eigenchannels for (a,b) majority and (c,d) minority
	spin components in FM arrangement. Blue and red indicate the positive and negative
	sign of the real part of the wavefunction.}
   \label{fig:fig3}
\end{figure}

To better understand the nature of spin transport in the system, we have furthermore calculated the spatially-resolved
scattering states in the contact region\cite{Paulsson2007}. The results are shown in~\ref{fig:fig3}.
Here, we consider the conducting FM arrangement, and focus on the two eigenchannel scattering states with highest transmission at $E_F$(moving in the direction up-to-down), which both are almost fully transmitting.
For the majority spins, we notice the $d_{zx}$ and $d_{yz}$ orbital nature of wavefunctions on the V adatoms
contacting \cs ($z$ chosen perpendicular to the surface).
This is in accordance with the Mulliken population analysis of the majority spin states of the V tip and adatoms,
where the $d_{zx}$, $d_{yz}$ each appears half-filled. On the other hand, the $d_{xy}, d_{x^2-y^2}$ and $d_z^2$
are closer to being filled, while the $s$ is closer to being empty. This points to a charge transfer from the V atoms to the \cs leaving the $d_{zx}/d_{yz}$ orbital energies closest to $E_F$. Since the $d_{zx}/d_{yz}$ orbitals match the symmetry with angular momentum $m=1$ for rotation around the V-\cs-V axis of the pentagon-prone
3-fold degenerate LUMO states (t$_{1u}$ symmetry \cite{HABRRA.1986,HaDuBa.2010}), we can expect the observed orbitals in the transport channels. For minority spins $d_{zx}$, $d_{yz}$ orbitals are almost empty and shifted away from $E_F$, resulting in a vanishing transmission. The rotational symmetric $m=0$ channels appear as resonances in the channel transmissions above $E_F$ and thus play a minor role.

\begin{figure}[htbp]
   \includegraphics[width=0.7\columnwidth]{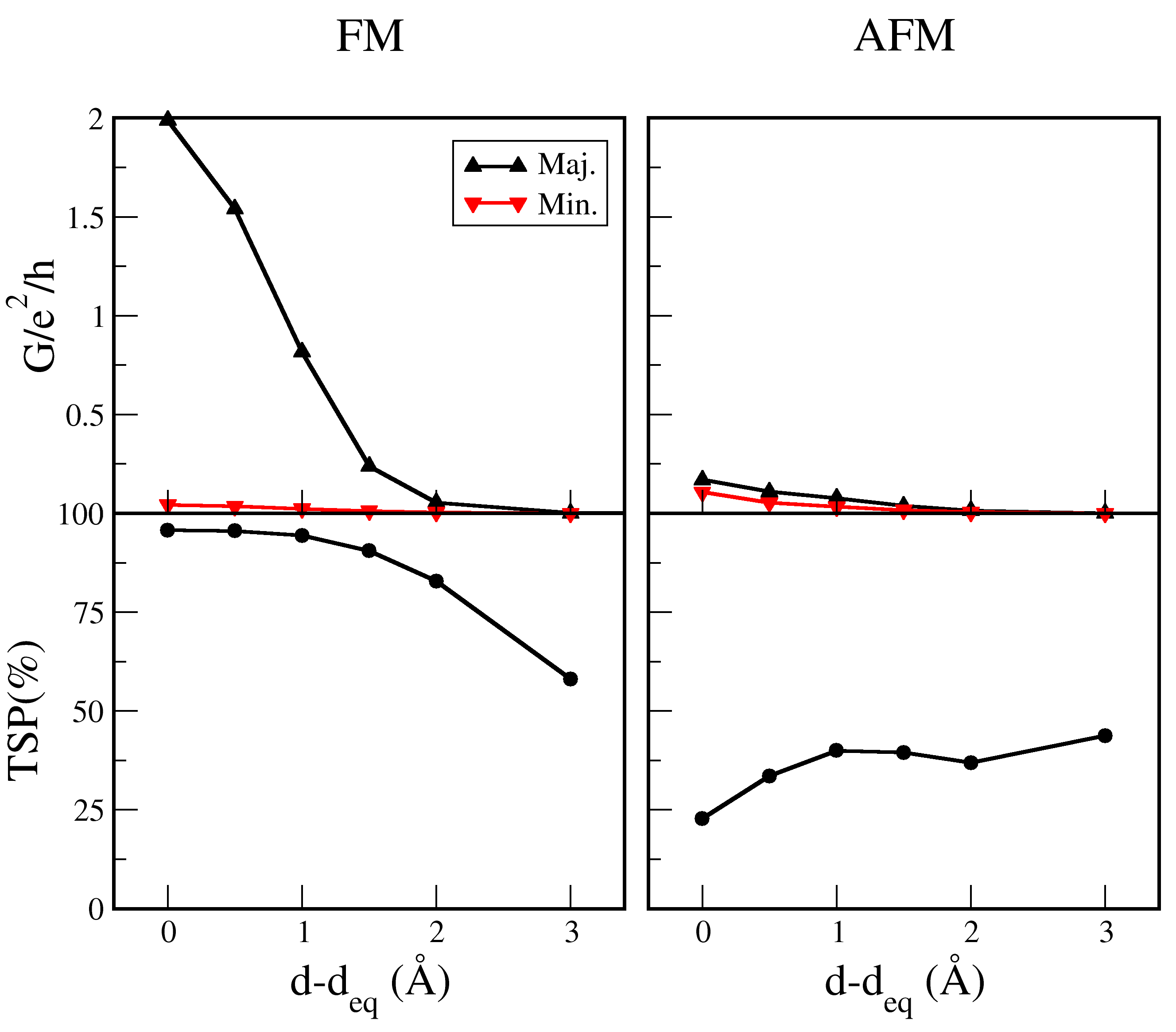}
   \caption{Spin-resolved conductance and transmission spin polarization(TSP) vs. \cs-adatom separation.}
   \label{fig:fig4}
\end{figure}

In typical STM experiments the conductance is probed from the tunnel-regime to contact.
We have performed transport calculations as the tip is approaching the surface adatom, until the 
tip-molecule distance ($d$ shown in~\ref{fig:fig1}) approximately reaches the equilibrium distance discussed above.
In~\ref{fig:fig4} we display the conductance along with the corresponding TSP as a function of tip distance. One can observe
the trend of increasing the conductance of the majority spins and thus TSP, in the FM case, while in AFM, the
conductance values are considerably smaller all the way to the equilibrium contact distance.

\begin{figure}[htbp]
   \includegraphics[width=0.7\columnwidth]{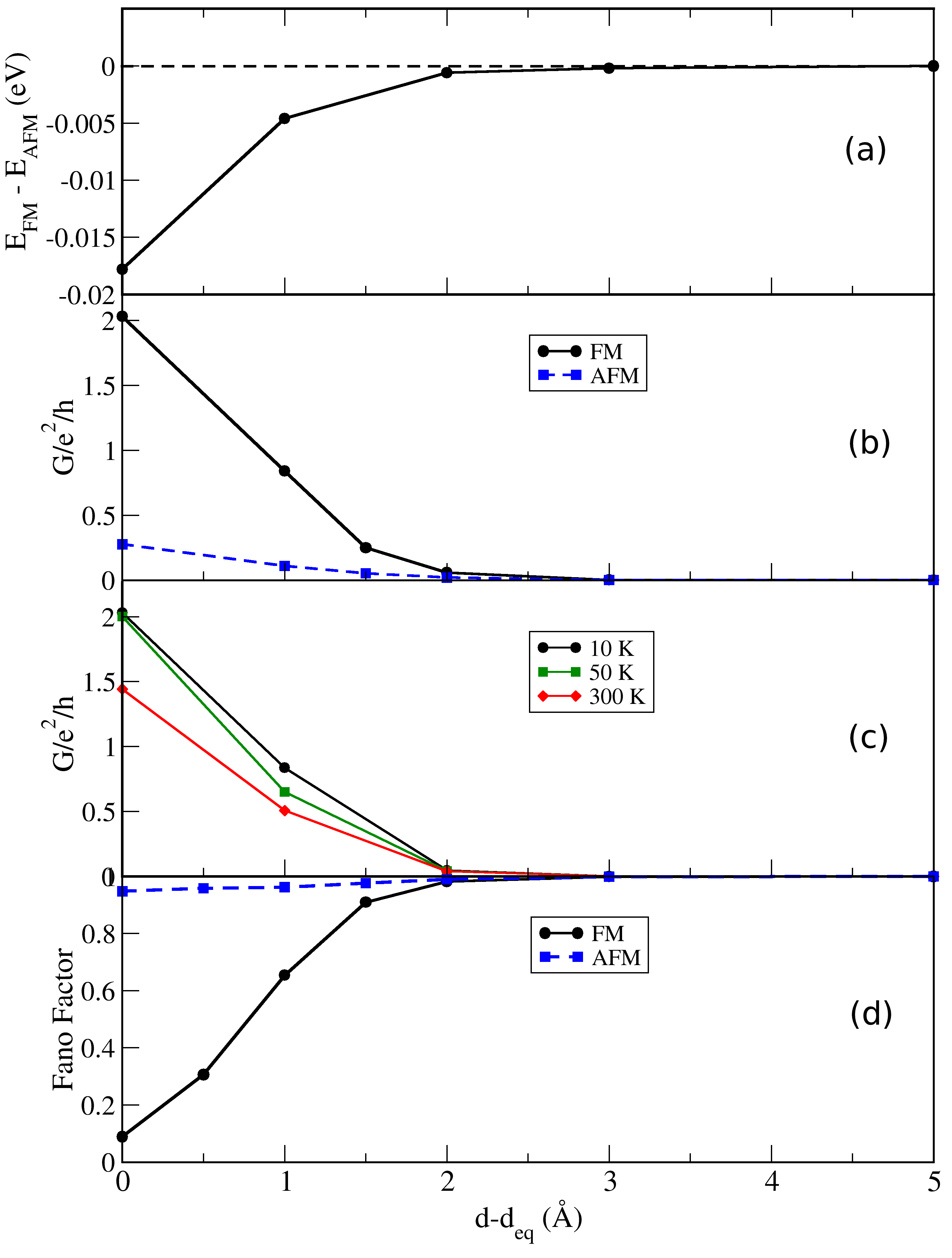}
   \caption{(a) magnetic exchange energy, (b) conductance for FM and AFM configurations, (c) thermal average of the conductance for different temperatures,
	(d) Fano factor of transmission	as a function of \cs- adatom separation for the FM and AFM configurations.}
   \label{fig:fig5}
\end{figure}

The marked difference between the FM and AFM conductance properties indicates how it is possible to probe the spin-coupling mediated by the \cs between the magnetic tip and substrate. The calculated magnetic interaction between the tip and adatom, the magnetic exchange energy defined as $E_{FM}-E_{AFM}$, is shown in~\ref{fig:fig5}(a) when the tip-molecule is approaching the adatom. This shows that the FM arrangement becomes favorable as the molecule reaches the equilibrium distance to the surface adatom. To be sure about the fidelity of the values obtained here, we have performed
the same study using the PW method. We find that the trend is the same and the values are even more pronounced in favor of FM arrangement, though of the same order of magnitude.

A graph of the total conductance versus \cs-adatom distance is shown in~\ref{fig:fig5}(b). The conductance behavior demonstrates a \emph{magnetic} valve, being closed for FM and open for AFM, if we imagine an external control over the magnetization of tip/substrate. In a typical experiment with a bulk magnetic tip the magnetization of the tip will be determined by the intrinsic magnetic anisotropy of the crystalline magnetization, which fix the magnetization axes. As the tip-molecule approaches the adatom on the non-magnetic surface, the magnetization of this will be determined by the interaction with the tip mediated by the molecule. In this case the adatom magnetization will align according to the thermal occupations. Using the calculated energies for the FM/AFM configurations we can now calculate the conductance for a given temperature for each distance,
\begin{equation}
G(d,T)=\frac{e^{-E_{FM}(d)/k_B T}G_{FM}(d)+e^{-E_{AFM}(d)/k_B T}G_{AFM}(d)}{e^{-E_{FM}(d)/k_B T}+e^{-E_{AFM}(d)/k_B T}}\,.
\end{equation}
This is shown in~\ref{fig:fig5}(c) for temperatures of 10K, 50K and 300K, showing only the FM configuration at low temperature.

The absolute distance is typically not known in an actual STM experiment. In principle, a particular conductance could be realized with both FM or AFM spin configurations -- a conductance of $e^2/h$ could result from a single spin-channel with perfect transmission or two half-transmitting channels. In combination with measurements of the conductance,
measurements of current shot-noise as characterized by the Fano factor,
\begin{equation}\label{eq.fano}
F = \frac{\sum_n T_n(E_F)(1-T_n(E_F))}{\sum_n T_n(E_F)}\,,
\end{equation}
 can provide further insights into the distribution of transmissions in the conductance channels, as demonstrated for molecular contacts\cite{DjRu.2006,KiTaWo.2008}.
In~\ref{fig:fig5}(d), we observe how the noise is significantly smaller for the FM configuration and drops already well before contact ($d-d_{eq}\sim 1.5${\AA}) is established. It can also be inferred how the transport is carried by two channels in the case of FM in contact.

In summary, we have shown how \cs in a STM junction formed between a V tip and a V adatom on a Cu substrate display highly
spin-polarized transport and a remarkable difference in conductance and noise properties depending on the
spin-alignment of the tip and adatom. Our results illustrate how the unique manipulation capabilities of the STM and the
high symmetry of the \cs may be exploited in order to gain insights into spin-transport and magnetic interactions at the molecular scale.

\acknowledgement
We thank Dr. Thomas Frederiksen for fruitful discussions.
M.K. acknowledges the financial support from land of Bremen.

\providecommand*{\mcitethebibliography}{\thebibliography}
\csname @ifundefined\endcsname{endmcitethebibliography}
{\let\endmcitethebibliography\endthebibliography}{}


\begin{mcitethebibliography}{34}
\providecommand*{\natexlab}[1]{#1}
\providecommand*{\mciteSetBstSublistMode}[1]{}
\providecommand*{\mciteSetBstMaxWidthForm}[2]{}
\providecommand*{\mciteBstWouldAddEndPuncttrue}
  {\def\EndOfBibitem{\unskip.}}
\providecommand*{\mciteBstWouldAddEndPunctfalse}
  {\let\EndOfBibitem\relax}
\providecommand*{\mciteSetBstMidEndSepPunct}[3]{}
\providecommand*{\mciteSetBstSublistLabelBeginEnd}[3]{}
\providecommand*{\EndOfBibitem}{}
\mciteSetBstSublistMode{f}
\mciteSetBstMaxWidthForm{subitem}{(\alph{mcitesubitemcount})}
\mciteSetBstSublistLabelBeginEnd{\mcitemaxwidthsubitemform\space}
{\relax}{\relax}

\bibitem[Xiong et~al.({2004})Xiong, Wu, Vardeny, and Shi]{XiWuVa.2004}
Xiong,~Z.; Wu,~D.; Vardeny,~Z.; Shi,~J. \emph{{NATURE}} \textbf{{2004}},
  \emph{{427}}, {821--824}\relax
\mciteBstWouldAddEndPuncttrue
\mciteSetBstMidEndSepPunct{\mcitedefaultmidpunct}
{\mcitedefaultendpunct}{\mcitedefaultseppunct}\relax
\EndOfBibitem
\bibitem[Rocha et~al.({2005})Rocha, Garcia-Suarez, Bailey, Lambert, Ferrer, and
  Sanvito]{RoGaBa.2005}
Rocha,~A.; Garcia-Suarez,~V.; Bailey,~S.; Lambert,~C.; Ferrer,~J.; Sanvito,~S.
  \emph{{NATURE MATERIALS}} \textbf{{2005}}, \emph{{4}}, {335--339}\relax
\mciteBstWouldAddEndPuncttrue
\mciteSetBstMidEndSepPunct{\mcitedefaultmidpunct}
{\mcitedefaultendpunct}{\mcitedefaultseppunct}\relax
\EndOfBibitem
\bibitem[Koopmans et~al.({2011})Koopmans, Wagemans, Bloom, Bobbert, Kemerink,
  and Wohlgenannt]{KoWaBl.2011}
Koopmans,~B.; Wagemans,~W.; Bloom,~F.~L.; Bobbert,~P.~A.; Kemerink,~M.;
  Wohlgenannt,~M. \emph{{PHILOSOPHICAL TRANSACTIONS OF THE ROYAL SOCIETY
  A-MATHEMATICAL PHYSICAL AND ENGINEERING SCIENCES}} \textbf{{2011}},
  \emph{{369}}, {3602--3616}\relax
\mciteBstWouldAddEndPuncttrue
\mciteSetBstMidEndSepPunct{\mcitedefaultmidpunct}
{\mcitedefaultendpunct}{\mcitedefaultseppunct}\relax
\EndOfBibitem
\bibitem[Herrmann et~al.({2010})Herrmann, Solomon, and
  Ratner]{ISI:000275868700022}
Herrmann,~C.; Solomon,~G.~C.; Ratner,~M.~A. \emph{{JOURNAL OF THE AMERICAN
  CHEMICAL SOCIETY}} \textbf{{2010}}, \emph{{132}}, {3682+}\relax
\mciteBstWouldAddEndPuncttrue
\mciteSetBstMidEndSepPunct{\mcitedefaultmidpunct}
{\mcitedefaultendpunct}{\mcitedefaultseppunct}\relax
\EndOfBibitem
\bibitem[Schull et~al.({2009})Schull, Frederiksen, Brandbyge, and
  Berndt]{ScFrBr.2009}
Schull,~G.; Frederiksen,~T.; Brandbyge,~M.; Berndt,~R. \emph{{PHYSICAL REVIEW
  LETTERS}} \textbf{{2009}}, \emph{{103}}, year\relax
\mciteBstWouldAddEndPuncttrue
\mciteSetBstMidEndSepPunct{\mcitedefaultmidpunct}
{\mcitedefaultendpunct}{\mcitedefaultseppunct}\relax
\EndOfBibitem
\bibitem[Berndt et~al.({2010})Berndt, Kroeger, Neel, and Schull]{BeKrNe.2010a}
Berndt,~R.; Kroeger,~J.; Neel,~N.; Schull,~G. \emph{{PHYSICAL CHEMISTRY
  CHEMICAL PHYSICS}} \textbf{{2010}}, \emph{{12}}, {1022--1032}\relax
\mciteBstWouldAddEndPuncttrue
\mciteSetBstMidEndSepPunct{\mcitedefaultmidpunct}
{\mcitedefaultendpunct}{\mcitedefaultseppunct}\relax
\EndOfBibitem
\bibitem[Schull et~al.({2011})Schull, Frederiksen, Arnau, Sanchez-Portal, and
  Berndt]{ScFrAr.2011}
Schull,~G.; Frederiksen,~T.; Arnau,~A.; Sanchez-Portal,~D.; Berndt,~R.
  \emph{{NATURE NANOTECHNOLOGY}} \textbf{{2011}}, \emph{{6}}, {23--27}\relax
\mciteBstWouldAddEndPuncttrue
\mciteSetBstMidEndSepPunct{\mcitedefaultmidpunct}
{\mcitedefaultendpunct}{\mcitedefaultseppunct}\relax
\EndOfBibitem
\bibitem[Martin et~al.(2008)Martin, Ding, S{\o}rensen, Bj{\o}rnholm, van
  Ruitenbeek, and van~der Zant]{martin2008fullerene}
Martin,~C.; Ding,~D.; S{\o}rensen,~J.; Bj{\o}rnholm,~T.; van Ruitenbeek,~J.;
  van~der Zant,~H. \emph{Journal of the American Chemical Society}
  \textbf{2008}, \emph{130}, 13198--13199\relax
\mciteBstWouldAddEndPuncttrue
\mciteSetBstMidEndSepPunct{\mcitedefaultmidpunct}
{\mcitedefaultendpunct}{\mcitedefaultseppunct}\relax
\EndOfBibitem
\bibitem[Braun et~al.(2009)Braun, Salaneck, and Fahlman]{braun2009energy}
Braun,~S.; Salaneck,~W.; Fahlman,~M. \emph{Advanced Materials} \textbf{2009},
  \emph{21}, 1450--1472\relax
\mciteBstWouldAddEndPuncttrue
\mciteSetBstMidEndSepPunct{\mcitedefaultmidpunct}
{\mcitedefaultendpunct}{\mcitedefaultseppunct}\relax
\EndOfBibitem
\bibitem[Bode({2003})]{Bo.2003}
Bode,~M. \emph{{REPORTS ON PROGRESS IN PHYSICS}} \textbf{{2003}}, \emph{{66}},
  {523--582}\relax
\mciteBstWouldAddEndPuncttrue
\mciteSetBstMidEndSepPunct{\mcitedefaultmidpunct}
{\mcitedefaultendpunct}{\mcitedefaultseppunct}\relax
\EndOfBibitem
\bibitem[Wiesendanger et~al.(1990)Wiesendanger, G\"untherodt, G\"untherodt,
  Gambino, and Ruf]{PhysRevLett.65.247}
Wiesendanger,~R.; G\"untherodt,~H.-J.; G\"untherodt,~G.; Gambino,~R.~J.;
  Ruf,~R. \emph{Phys. Rev. Lett.} \textbf{1990}, \emph{65}, 247--250\relax
\mciteBstWouldAddEndPuncttrue
\mciteSetBstMidEndSepPunct{\mcitedefaultmidpunct}
{\mcitedefaultendpunct}{\mcitedefaultseppunct}\relax
\EndOfBibitem
\bibitem[Hirjibehedin et~al.(2006)Hirjibehedin, Lutz, and
  Heinrich]{hirjibehedin2006spin}
Hirjibehedin,~C.; Lutz,~C.; Heinrich,~A. \emph{Science} \textbf{2006},
  \emph{312}, 1021\relax
\mciteBstWouldAddEndPuncttrue
\mciteSetBstMidEndSepPunct{\mcitedefaultmidpunct}
{\mcitedefaultendpunct}{\mcitedefaultseppunct}\relax
\EndOfBibitem
\bibitem[Neel et~al.({2010})Neel, Kroeger, and Berndt]{ISI:000286768600002}
Neel,~N.; Kroeger,~J.; Berndt,~R. \emph{{PHYSICAL REVIEW B}} \textbf{{2010}},
  \emph{{82}}, year\relax
\mciteBstWouldAddEndPuncttrue
\mciteSetBstMidEndSepPunct{\mcitedefaultmidpunct}
{\mcitedefaultendpunct}{\mcitedefaultseppunct}\relax
\EndOfBibitem
\bibitem[Liljeroth et~al.({2010})Liljeroth, Swart, Paavilainen, Repp, and
  Meyer]{ISI:000280416200031}
Liljeroth,~P.; Swart,~I.; Paavilainen,~S.; Repp,~J.; Meyer,~G. \emph{{NANO
  LETTERS}} \textbf{{2010}}, \emph{{10}}, {2475--2479}\relax
\mciteBstWouldAddEndPuncttrue
\mciteSetBstMidEndSepPunct{\mcitedefaultmidpunct}
{\mcitedefaultendpunct}{\mcitedefaultseppunct}\relax
\EndOfBibitem
\bibitem[Tao et~al.(2009)Tao, Stepanyuk, Hergert, Rungger, Sanvito, and
  Bruno]{tao2009switching}
Tao,~K.; Stepanyuk,~V.; Hergert,~W.; Rungger,~I.; Sanvito,~S.; Bruno,~P.
  \emph{Physical review letters} \textbf{2009}, \emph{103}, 57202\relax
\mciteBstWouldAddEndPuncttrue
\mciteSetBstMidEndSepPunct{\mcitedefaultmidpunct}
{\mcitedefaultendpunct}{\mcitedefaultseppunct}\relax
\EndOfBibitem
\bibitem[Polok et~al.({2011})Polok, Fedorov, Bagrets, Zahn, and
  Mertig]{PoFeBa.2011}
Polok,~M.; Fedorov,~D.~V.; Bagrets,~A.; Zahn,~P.; Mertig,~I. \emph{{PHYSICAL
  REVIEW B}} \textbf{{2011}}, \emph{{83}}, year\relax
\mciteBstWouldAddEndPuncttrue
\mciteSetBstMidEndSepPunct{\mcitedefaultmidpunct}
{\mcitedefaultendpunct}{\mcitedefaultseppunct}\relax
\EndOfBibitem
\bibitem[Brune and Gambardella({2009})]{BrGa.2009}
Brune,~H.; Gambardella,~P. \emph{{SURFACE SCIENCE}} \textbf{{2009}},
  \emph{{603}}, {1812--1830}\relax
\mciteBstWouldAddEndPuncttrue
\mciteSetBstMidEndSepPunct{\mcitedefaultmidpunct}
{\mcitedefaultendpunct}{\mcitedefaultseppunct}\relax
\EndOfBibitem
\bibitem[Atodiresei et~al.(2010)Atodiresei, Brede, Lazi{\'c}, Caciuc, Hoffmann,
  Wiesendanger, and Bl{\"u}gel]{atodiresei2010design}
Atodiresei,~N.; Brede,~J.; Lazi{\'c},~P.; Caciuc,~V.; Hoffmann,~G.;
  Wiesendanger,~R.; Bl{\"u}gel,~S. \emph{Physical review letters}
  \textbf{2010}, \emph{105}, 66601\relax
\mciteBstWouldAddEndPuncttrue
\mciteSetBstMidEndSepPunct{\mcitedefaultmidpunct}
{\mcitedefaultendpunct}{\mcitedefaultseppunct}\relax
\EndOfBibitem
\bibitem[Crabtree(2009)]{crabtree2009organometallic}
Crabtree,~R. \emph{The organometallic chemistry of the transition metals};
\newblock Wiley New York etc., 2009\relax
\mciteBstWouldAddEndPuncttrue
\mciteSetBstMidEndSepPunct{\mcitedefaultmidpunct}
{\mcitedefaultendpunct}{\mcitedefaultseppunct}\relax
\EndOfBibitem
\bibitem[Tao et~al.(2008)Tao, Stepanyuk, Bruno, Bazhanov, Maslyuk, Brandbyge,
  and Mertig]{tao2008manipulating}
Tao,~K.; Stepanyuk,~V.; Bruno,~P.; Bazhanov,~D.; Maslyuk,~V.; Brandbyge,~M.;
  Mertig,~I. \emph{Physical Review B} \textbf{2008}, \emph{78}, 014426\relax
\mciteBstWouldAddEndPuncttrue
\mciteSetBstMidEndSepPunct{\mcitedefaultmidpunct}
{\mcitedefaultendpunct}{\mcitedefaultseppunct}\relax
\EndOfBibitem
\bibitem[Wang et~al.(2008)Wang, Cai, Wang, Lu, Luo, Lai, Zhou, Qin, Gao, Yu,
  Li, Mei, and Sanvito]{WaCaWa.2008}
Wang,~L.; Cai,~Z.; Wang,~J.; Lu,~J.; Luo,~G.; Lai,~L.; Zhou,~J.; Qin,~R.;
  Gao,~Z.; Yu,~D.; Li,~G.; Mei,~W.~N.; Sanvito,~S. \emph{Nano Letters}
  \textbf{2008}, \emph{8}, 3640--3644, PMID: 18816146\relax
\mciteBstWouldAddEndPuncttrue
\mciteSetBstMidEndSepPunct{\mcitedefaultmidpunct}
{\mcitedefaultendpunct}{\mcitedefaultseppunct}\relax
\EndOfBibitem
\bibitem[Maslyuk et~al.({2006})Maslyuk, Bagrets, Meded, Arnold, Evers,
  Brandbyge, Bredow, and Mertig]{MaBaMe.2006}
Maslyuk,~V.~V.; Bagrets,~A.; Meded,~V.; Arnold,~A.; Evers,~F.; Brandbyge,~M.;
  Bredow,~T.; Mertig,~I. \emph{{PHYSICAL REVIEW LETTERS}} \textbf{{2006}},
  \emph{{97}}, year\relax
\mciteBstWouldAddEndPuncttrue
\mciteSetBstMidEndSepPunct{\mcitedefaultmidpunct}
{\mcitedefaultendpunct}{\mcitedefaultseppunct}\relax
\EndOfBibitem
\bibitem[Koleini et~al.({2007})Koleini, Paulsson, and Brandbyge]{KoPaBr.2007}
Koleini,~M.; Paulsson,~M.; Brandbyge,~M. \emph{{PHYSICAL REVIEW LETTERS}}
  \textbf{{2007}}, \emph{{98}}, year\relax
\mciteBstWouldAddEndPuncttrue
\mciteSetBstMidEndSepPunct{\mcitedefaultmidpunct}
{\mcitedefaultendpunct}{\mcitedefaultseppunct}\relax
\EndOfBibitem
\bibitem[Weber et~al.(1997)Weber, Rao, Jena, Stepanyuk, Hergert, Wildberger,
  Zeller, and Dederichs]{weber1997magnetism}
Weber,~S.; Rao,~B.; Jena,~P.; Stepanyuk,~V.; Hergert,~W.; Wildberger,~K.;
  Zeller,~R.; Dederichs,~P. \emph{Journal of Physics: Condensed Matter}
  \textbf{1997}, \emph{9}, 10739\relax
\mciteBstWouldAddEndPuncttrue
\mciteSetBstMidEndSepPunct{\mcitedefaultmidpunct}
{\mcitedefaultendpunct}{\mcitedefaultseppunct}\relax
\EndOfBibitem
\bibitem[Soler et~al.(2002)Soler, Artacho, Gale, Garcia, Junquera, Ordejon, and
  Sanchez-Portal]{Soler2002}
Soler,~J.; Artacho,~E.; Gale,~J.; Garcia,~A.; Junquera,~J.; Ordejon,~P.;
  Sanchez-Portal,~D. \emph{J. Phys.: Condens. Mat.} \textbf{2002}, \emph{14},
  2745\relax
\mciteBstWouldAddEndPuncttrue
\mciteSetBstMidEndSepPunct{\mcitedefaultmidpunct}
{\mcitedefaultendpunct}{\mcitedefaultseppunct}\relax
\EndOfBibitem
\bibitem[Perdew et~al.(1996)Perdew, Burke, and Ernzerhof]{Perdew1996}
Perdew,~J.; Burke,~K.; Ernzerhof,~M. \emph{Phys. Rev. Lett.} \textbf{1996},
  \emph{77}, 3865--3868\relax
\mciteBstWouldAddEndPuncttrue
\mciteSetBstMidEndSepPunct{\mcitedefaultmidpunct}
{\mcitedefaultendpunct}{\mcitedefaultseppunct}\relax
\EndOfBibitem
\bibitem[Brandbyge et~al.(2002)Brandbyge, Mozos, Ordej{\'o}n, Taylor, and
  Stokbro]{Brandbyge2002}
Brandbyge,~M.; Mozos,~J.; Ordej{\'o}n,~P.; Taylor,~J.; Stokbro,~K. \emph{Phys.
  Rev. B} \textbf{2002}, \emph{65}, 165401\relax
\mciteBstWouldAddEndPuncttrue
\mciteSetBstMidEndSepPunct{\mcitedefaultmidpunct}
{\mcitedefaultendpunct}{\mcitedefaultseppunct}\relax
\EndOfBibitem
\bibitem[Giannozzi et~al.(2009)Giannozzi, Baroni, Bonini, Calandra, Car,
  Cavazzoni, Ceresoli, Chiarotti, Cococcioni, Dabo, Corso, de~Gironcoli,
  Fabris, Fratesi, Gebauer, Gerstmann, Gougoussis, Kokalj, Lazzeri,
  Martin-Samos, Marzari, Mauri, Mazzarello, Paolini, Pasquarello, Paulatto,
  Sbraccia, Scandolo, Sclauzero, Seitsonen, Smogunov, Umari, and
  Wentzcovitch]{Giannozzi2009}
Giannozzi,~P. et~al. \emph{J. Phys.: Condens. Matter} \textbf{2009}, \emph{21},
  395502 (19pp)\relax
\mciteBstWouldAddEndPuncttrue
\mciteSetBstMidEndSepPunct{\mcitedefaultmidpunct}
{\mcitedefaultendpunct}{\mcitedefaultseppunct}\relax
\EndOfBibitem
\bibitem[Paulsson and Brandbyge(2007)]{Paulsson2007}
Paulsson,~M.; Brandbyge,~M. \emph{Phys. Rev. B} \textbf{2007}, \emph{76},
  115117\relax
\mciteBstWouldAddEndPuncttrue
\mciteSetBstMidEndSepPunct{\mcitedefaultmidpunct}
{\mcitedefaultendpunct}{\mcitedefaultseppunct}\relax
\EndOfBibitem
\bibitem[HADDON et~al.({1986})HADDON, BRUS, and RAGHAVACHARI]{HABRRA.1986}
HADDON,~R.; BRUS,~L.; RAGHAVACHARI,~K. \emph{{CHEMICAL PHYSICS LETTERS}}
  \textbf{{1986}}, \emph{{125}}, {459--464}\relax
\mciteBstWouldAddEndPuncttrue
\mciteSetBstMidEndSepPunct{\mcitedefaultmidpunct}
{\mcitedefaultendpunct}{\mcitedefaultseppunct}\relax
\EndOfBibitem
\bibitem[Hands et~al.({2010})Hands, Dunn, and Bates]{HaDuBa.2010}
Hands,~I.~D.; Dunn,~J.~L.; Bates,~C.~A. \emph{{PHYSICAL REVIEW B}}
  \textbf{{2010}}, \emph{{81}}, year\relax
\mciteBstWouldAddEndPuncttrue
\mciteSetBstMidEndSepPunct{\mcitedefaultmidpunct}
{\mcitedefaultendpunct}{\mcitedefaultseppunct}\relax
\EndOfBibitem
\bibitem[Djukic and van Ruitenbeek({2006})]{DjRu.2006}
Djukic,~D.; van Ruitenbeek,~J. \emph{{NANO LETTERS}} \textbf{{2006}},
  \emph{{6}}, {789--793}\relax
\mciteBstWouldAddEndPuncttrue
\mciteSetBstMidEndSepPunct{\mcitedefaultmidpunct}
{\mcitedefaultendpunct}{\mcitedefaultseppunct}\relax
\EndOfBibitem
\bibitem[Kiguchi et~al.(2008)Kiguchi, Tal, Wohlthat, Pauly, Krieger, Djukic,
  Cuevas, and van Ruitenbeek]{KiTaWo.2008}
Kiguchi,~M.; Tal,~O.; Wohlthat,~S.; Pauly,~F.; Krieger,~M.; Djukic,~D.;
  Cuevas,~J.~C.; van Ruitenbeek,~J.~M. \emph{Phys. Rev. Lett.} \textbf{2008},
  \emph{101}, 046801\relax
\mciteBstWouldAddEndPuncttrue
\mciteSetBstMidEndSepPunct{\mcitedefaultmidpunct}
{\mcitedefaultendpunct}{\mcitedefaultseppunct}\relax
\EndOfBibitem
\end{mcitethebibliography}
\end{document}